\documentstyle[aps,preprint]{revtex}
\begin{document}
\title{Spinorial density matrix equation and gauge covariance}
\author{F. C. Khanna${}^{a,b}$, A. E. Santana{}$^c$, A. Matos Neto${}^c$,}
\author{J. D. M. Vianna${}^{c,d}$ and T. Kopf\thinspace $^a$}
\address{${}^a$Theoretical Physics Institute, Dept. of Physics, \\
Univiversity of Alberta, \\
Edmonton, AB T6G 2J1, Canada;}
\address{${}^b$TRIUMF, 4004, Wesbrook Mall, \\
V6T 2A3, Vancouver, BC, Canada;}
\address{${}^c$Instituto de F\'\i sica, Universidade Federal da Bahia, \\
Campus de Ondina, 40210-340, Salvador, Bahia, Brasil; }
\address{${}^d$Departamento de F\'\i sica, Universidade de Bras\'\i lia, 70910-900,\\
Bras\'\i lia, DF, Brasil.}
\maketitle

\begin{abstract}
In this work we apply the Lie group representation method introduced in the
real time formalism for finite-temperature quantum-field theory, thermofield
dynamics, to derive a spinorial density matrix equation. Symmetry properties
of such equation are analysed, and as a basic result it is shown that one
solution is the generalised density matrix operator proposed by Heinz, to
deal with gauge covariant kinetic equations. In the same context,
preliminary aspects of a Lagrangian formalism to derive kinetic equations,
as well as quantum density matrix equations in curved space-time, are
discussed.

\thinspace \thinspace \thinspace \thinspace \thinspace \thinspace \thinspace
\thinspace \thinspace \thinspace \thinspace \thinspace \thinspace \thinspace 
${\rm \,manuscript,17pages}$
\end{abstract}

\section{Introduction}

This paper presents a group symmetry scheme in order to treat the
relatisvistic density matrix including the gauge covariance. In a previous
paper\cite{ak1} (to be referred from now on as {\it I}), we have set forth a
representation theory for Lie groups, which is based in the thermofield
dynamics (TFD), a real-time finite-temperature field theory via an operator
method \cite{ume1,ume2,ume3}. In {\it I} the TFD Hilbert space has been used
as the representation space for Lie algebras, and so, the equations of
motion in quantum TFD have been derived from the Galilei and Poincar\'e
symmetries\cite{ak1,ak2}; a classical counterpart of TFD has been identified%
\cite{ak3,ak4}; and a connection of TFD with the GNS (Gel'fand, Naimark and
Segal) construction\cite{gns} has been analysed in particular in association
with elements of Hopf algebras\cite{ak2,ak5,ak6,ko1,ak7}. A density-matrix
equation for the Klein-Gordon field has been introduced and basic aspects of
the kinetic theory for bosons has been analysed\cite{ak2}. This has
indicated the possibility to extend this symmetry group formalism for other
systems of the quantum kinetic field theory, in particular by considering
gauge fields and curved space-time. Such a study can bring the relativistic
transport equations to a more fundamental level of symmetry group, and such
aspect find interest in the recent developments in high energy physics and
in early universe theories.

One remarkable prediction of the quantum chromodynamics is the possibility
of a transition from a confined hadronic matter to a deconfined quark-gluon
plasma (QGP), in an ultra-relativistic collision of heavy ions, at an energy
density of ${\sim }2\,\,GeV/fm^3$ and a temperature of $\sim 200MeV$.
Experimentally, such conditions will be realized soon, and this fact makes
it essential to elaborate on some aspects, such as the signature of the QGP
in the pre-equilibrium stage\cite{q1}. These developments have provided a
strong motivation for a revival of the kinetic theory in order to treat
classical or quantum quark-gluon systems\cite{q1,q2}. It is our contention
that the symmetry properties can provide an important tool for such
developments. Therefore, this paper explores the fundamental consequences of
symmetry on the kinetic theory under gauge invariance as in the case of a
quark-gluon plasma.

In order to deal with gauge kinetic-theory models, Heinz\cite
{hein1,hein2,hein3,hein4,hein5,hein6,hein7,hein8} has proposed a
generalization of the concept of density matrix to take into account
explicitly the gauge fields. Considering a field $\phi (x)$, Heinz has
written the density matrix in the following form 
\begin{equation}
\Psi (x,y)=e^{{-yd(x)}}\ \phi (x)\otimes \bar \phi (x)\ e^{{yd(x)^{\dagger }}%
},  \label{hee1}
\end{equation}
where ${d(x)}$ is a covariant derivative. Such a theory has been successful
in introducing gauge invariant Wigner operators in relativistic phase
spaces, such that transport equations have been derived, though, the
collision terms have not been treated in a full and consistent way (indeed,
difficulties with the collision terms are a characteristic aspect of the
present state of the art of the gauge kinetic theory). One way to proceed in
order to derive a full-collision kinetic approach is, first of all, to
understand the underlying structure attached to the Heinz density operator.
One such answer is presented in this paper, where we derive (\ref{hee1}) as
a solution of a density matrix equation proposed here via the representation
theory of {\it I }considering the spinorial field. This result suggests a
generalization of the Heinz density operator (\ref{hee1}) to deal with
quantum fields in curved space-time.

An approach to derive transport equations for systems treated by a quantum
field in curved space-time background was proposed by Calzetta, Habib and Hu%
\cite{ca1} and Calzetta and Hu\cite{ca2}. In such a method problems related
to the notion of particles in curved manifold, averages, distribution
functions and distance between two points, all of them necessary to build a
kinetic theory, have been partially overcome through the use of the Riemann
normal coordinates, in which, in the neighborhood of the origin of
coordinates, some modes are close to plane waves. So the metric tensor $%
g_{\mu \nu }$ is expanded as 
\[
g_{\mu \nu }(x)=\eta _{\mu \nu }+\frac 13R_{\mu \tau \nu \rho }x^\tau x^\rho
+..., 
\]
where $\eta _{\mu \nu }$ is a flat-space metric and $R_{\mu \tau \nu \rho }$
is the Riemann tensor. On the other hand , information about early universes
may be obtained by the time evolution of the density matrix of the universe,
by considering the interaction of the quantum fields and many-particles
systems with the classical gravitational field. An important problem in this
context is of the gauge quantum kinetic theory, which we are interested to
consider at this stage.

We proceed in the following way. In section 2 some structural elements of
TFD are presented in connection with the representation theory developed in 
{\it I}, but rederiving the basic method along with our proposals here. In
section 3 the density matrix equation is introduced via a representation of
the Poincar\'e group and the gauge invariance is analysed in this context.
In section 4 we derive a collisionless Boltzmann equation for the spinorial
field in an algebraic way, such that the approach proposed here can be
better compared with others in the literature. A possible Lagrangian
formalism for the kinetic theory is discussed and some comments are made
about the extension to the curved space-time as suggested by Calzetta, Habib
and Hu\cite{ca1,ca2}. Final remarks and conclusions are presented in Section
5.

\section{Thermofield dynamics and representation theory for Lie groups}

The thermofield dynamics is based on two main ingredients: one , the dual
(or tilde) conjugation rules, introduces a doubling in the dynamical
variable; while the other, the Bogolioubov transformation, introduces
thermal effects through a vacuum condensation process. The doubling, which
is a very attractive algebraic aspect in TFD, is defined in the following
way.

Let ${\cal L}=\left\{ A_i,i=1,2,3,...n;a,b,c,...\in {\cal C}\right\} $ be an
associative algebra on a ${\cal C}$-field. Tilde conjugation is a mapping ${%
\tau }$ : ${\cal L}{\rightarrow }\ {\cal L}$, denoted as $\tau A\tau ={%
\tilde A}$, furnishing the following properties\cite{ume3}: 
\begin{eqnarray}
{(A_iA_j{\tilde )}} &=&{\tilde A}_i{\tilde A}_j,  \label{til} \\
(cA_i+A_j{\tilde )} &=&{c^{*}}{\tilde A}_i+{\tilde A}_j,  \label{til2} \\
(A_i^{\dag }{\tilde )} &=&{({\tilde A}_i)}^{\dag }, \\
({\tilde A}_i{\tilde )} &=&A_i, \\
{\lbrack A_i,{\tilde A}_j]} &=&0,  \label{til5}
\end{eqnarray}
where ${[\ ,\ ]}$ is the commutator. Usually, in TFD, there are two more
rules: $|0(\theta )\tilde \rangle =|0(\theta )\rangle $ and $\tilde \langle
0(\theta )|=\langle 0(\theta )|,$ which show that vacua in TFD are
tilde-conjugation invariant\cite{ume3}.

Automorphisms in ${\cal L}$ can then be introduced through unitary
operators, say $U(\xi )$, such that we require $\tau U(\xi )\tau =U(\xi )$.
With this result, writing $U(\xi )$ as $U(\xi )=\exp (i\xi \widehat{A})$,
where $\widehat{A}$ is a transformation generator, we have $\tau \widehat{A}%
\tau =-\widehat{A}$. Therefore, $\widehat{A}$ can be considered as an odd
polynomial function of $A-\widetilde{A}$, i.e., 
\begin{equation}
\widehat{A}=f(A-\widetilde{A})=\sum_{n=0}^\infty c_n(A-\widetilde{A})^{2n+1},
\label{um}
\end{equation}
where the coefficients $c_n$ are c-numbers.

We consider here the simplest case with $n=0$, that is, 
\begin{equation}
\widehat{A}=A-\widetilde{A}.  \label{dois}
\end{equation}
Therefore, this, Eq.(\ref{dois}), equips the TFD algebra, ${\cal L}$, with
an hat-isomorphism, satisfying the properties:

\begin{eqnarray}
(cA+B)\,\widehat{} &=&c\,\widehat{A}+\widehat{B},\,\,\,c\in {\bf R,}
\label{p1} \\
(icA)\,\widehat{} &=&ic(2A-\widehat{A}),\,  \label{p2} \\
(AB)\,\widehat{} &=&A\,\widehat{B}+\,\widehat{A}B-\widehat{A}\widehat{B},\,\,
\label{p3} \\
\lbrack A,B]\,\widehat{} &=&[A,\,\widehat{B}]+[\,\widehat{A},B]-[\widehat{A},%
\widehat{B}],  \label{p4} \\
\lbrack A,B]_{+}\,\widehat{} &=&[A,\,\widehat{B}]_{+}+[\,\widehat{A},B]_{+}-[%
\widehat{A},\widehat{B}]_{+},  \label{p5} \\
\widehat{\widehat{A}} &=&2\widehat{A}.  \label{p6}
\end{eqnarray}
The proof follows from the definition of tilde conjugation rules and Eq.(\ref
{dois}).

Now we are going to explore these generators ($\widehat{A}$), as elements of
a Lie algebra. Consider $\ell =\{a_i,i=1,2,3,...\}$ a Lie algebra over the
(real) field ${\bf R}$, characterized by the algebraic relations $%
a_i\bigtriangleup a_j=C_{ij}^ka_k$, where $C_{ij}^k\in {\bf R}$ are the
structure constants and $\bigtriangleup $ is the Lie product (we are using
the convention of sum over repeated indices). Taking the TFD Hilbert space,
say ${\cal H}_w$, as the space of representations for $\ell $, we write

\begin{eqnarray}
\lbrack \widehat{A}_i,\widehat{A}_j]&=&iC_{ij}^k\widehat{A}_k.  \label{ter1}
\\
\lbrack \widehat{A}_i,A_j] &=&iD_{ij}^kA_k,  \label{ter2} \\
\lbrack A_i,A_j] &=&iE_{ij}^kA_k,.  \label{ter3}
\end{eqnarray}

That is to say, physically Eq.(\ref{ter1}) represents the Lie symmetries
characterized by the structure constants $C_{ij}^k$ (by definition, the hat
operators $\widehat{A}$ are the symmetry generators, a class of dynamical
observables). Since the non-hat operators $(A)$ can be identified with the
observables, the other possible class of dynamical variables, then Eq.(\ref
{ter2}) shows the way that generators of symmetries determine dynamical
changes in the observables. The Abelian (or non-Abelian) nature of the
observables in regard to the measurement process is establised by Eq.(\ref
{ter3}) (this interpretation has been used to write Eq.(\ref{ter3}) as $%
[A_i,A_j]=0$ and so to derive a classical TFD structure\cite{ak3,ak4}).

In this procedure, we have split the twofold structure of ${\cal H}_w$, by
introducing a representation for Lie symmetry, defined by the (TFD) algebra
given by Eqs.(\ref{ter1})-(\ref{ter3}), which we denote by $^{*}{\ell }$
(regarding the close connection of this structure with the w$^{*}$-algebra,
see Ref. \cite{ak2}). In the next section we use this representation theory
to analyse a spinorial representation for the $^{*}$-Poincar\'e group, $%
^{*}p $.

Closing this section, let us observe that for a gauge symmetry, a set of
generators of the gauge group will be added to the space-time symmetry
group; and, of course, these two set of generators commute with each other.
However, the implications of gauge groups as additional quantum numbers,
such as color, etc, will apear in defining the Hilbert space.

\section{Poincar\'e group and density-matrix field equations}

Here we build a $^{*}$-$p$ Lie algebra assuming that in Eqs.(\ref{ter1})-(%
\ref{ter3}) we have $D_{ij}^k=E_{ij}^k=C_{ij}^k$, where $C_{ij}^k$ are the
structure constants of the Poincar\'e Lie algebra, such that we can write

\begin{eqnarray}
\lbrack M_{\mu \nu },P_\sigma ] &=&i(g_{\nu \sigma }P_\mu -g_{\sigma \mu
}P_\nu ),  \label{poin1} \\
\lbrack P_\mu ,P_\nu ] &=&0,  \label{poin2} \\
\lbrack M_{\mu \nu },M_{\sigma \rho }] &=&-i(g_{\mu \rho }M_{\nu \sigma
}-g_{\nu \rho }M_{\mu \sigma }+g_{\mu \sigma }M_{\rho \nu }-g_{\nu \sigma
}M_{\rho \mu }),  \label{poin3} \\
\lbrack \widehat{M}_{\mu \nu },P_\sigma ] &=&[M_{\mu \nu },\widehat{P}%
_\sigma ]=i(g_{\nu \sigma }P_\mu -g_{\sigma \mu }P_\nu ),  \label{poin4} \\
\lbrack \widehat{P}_\mu ,P_\nu ] &=&0,  \label{poin5} \\
\lbrack \widehat{M}_{\mu \nu },M_{\sigma \rho }] &=&-i(g_{\mu \rho }M_{\nu
\sigma }-g_{\nu \rho }M_{\mu \sigma }+g_{\mu \sigma }M_{\rho \nu }-g_{\nu
\sigma }M_{\rho \mu }),  \label{poin6} \\
\lbrack \widehat{M}_{\mu \nu },\widehat{P}_\sigma ] &=&i(g_{\nu \sigma }%
\widehat{P}_\mu -g_{\sigma \mu }\widehat{P}_\nu ),  \label{poin7} \\
\lbrack \widehat{P}_\mu ,\widehat{P}_\nu ] &=&0,  \label{poin8} \\
\lbrack \widehat{M}_{\mu \nu },\widehat{M}_{\sigma \rho }] &=&-i(g_{\mu \rho
}\widehat{M}_{\nu \sigma }-g_{\nu \rho }\widehat{M}_{\mu \sigma }+g_{\mu
\sigma }\widehat{M}_{\rho \nu }-g_{\nu \sigma }\widehat{M}_{\rho \mu }),
\label{poin9}
\end{eqnarray}
where $\widehat {M}_{\mu \nu }$ stands for the generators of rotations and $%
\widehat {P}_\mu $ the generators of translations; whilst $M_{\mu \nu }$ and 
$P_\mu$ are, respectively, the corresponding observables for rotations and
translations. The metric tensor is such that $\,\,\,diag(g_{\mu \nu
})=(1,-1,-1,-1),$ and $\,\,g_{\mu \nu }=0$ \thinspace \thinspace for \ $\mu
\neq \nu ;\,\,\,\,\,\mu ,\,\,\nu =0,1,2,3$.

The invariants of $^{*}p$ are 
\begin{eqnarray}
W &=&w_\mu w^\mu ,  \label{inv1} \\
P^2 &=&P_\mu P^\mu ,  \label{inv2} \\
\widehat{W} &=&w_\mu w^\mu -\widetilde{w}_\mu \widetilde{w}^\mu =2\widehat{w}%
_\mu w^\mu -\widehat{w}_\mu \widehat{w}^\mu ,  \label{inv3} \\
\widehat{P^2} &=&(P^2)\,^{\widehat{}}=P_\mu P^\mu -\widetilde{P}_\mu 
\widetilde{P}^\mu=2\widehat{P}_\mu P^\mu -\widehat{P}_\mu \widehat{P}^\mu ;
\label{inv4}
\end{eqnarray}
where $w_\mu =\frac 12\varepsilon _{\mu \nu \rho \sigma }M^{\nu \sigma
}P^\rho $ are the Pauli-Lubanski matrices; $\varepsilon _{\mu \nu \rho
\sigma }$ is the Levi-Civita symbol; and 
\[
\widehat{w}_\mu =\frac 12\varepsilon _{\mu \nu \rho \sigma }\widehat{M}^{\nu
\sigma }P^\rho +\frac 12\varepsilon _{\mu \nu \rho \sigma }M^{\nu \sigma }%
\widehat{P}^\rho -\frac 12\varepsilon _{\mu \nu \rho \sigma }\widehat{M}%
^{\nu \sigma }\widehat{P}^\rho . 
\]
Notice that the tensor 
\[
\overline{w}_\mu =\frac 12\varepsilon _{\mu \nu \rho \sigma }\widehat{M}%
^{\nu \sigma }\widehat{P}^\rho 
\]
can be used to define the scalar $\overline{W}=\overline{w}_\mu \overline{w}%
^\mu $ , which is not an invariant of $^{*}p$ but rather of the subalgebra
(of $^{*}p$ ) given by Eqs.(\ref{poin7})-(\ref{poin9}).

%
%
%
%
%

Denoting the vectors of the representation space of ${\cal H}_w$ by $|\Psi
\rangle $ such that 
\begin{equation}
|\Psi \rangle =|\phi \rangle \otimes \langle \phi |=|\phi \rangle \otimes |%
\widetilde{\phi }\rangle ;  \label{est11}
\end{equation}
where $|\phi \rangle$ is an element of a Hilbert space ${\cal H}$, such that 
${\cal H}_w={\cal H}{\otimes }{\cal H}^{*}$; we also assume that $\langle
\phi |\phi \rangle =1$. The action of the operators $O$ and ${\widetilde{O}}$
in ${\cal H}_w$ is then

\begin{eqnarray}
O|\Psi\rangle &{\rightarrow}&[O\otimes 1]|\Psi \rangle  \nonumber \\
&=&[O\otimes 1]|\phi \rangle \otimes \langle \phi |=O|\phi \rangle \otimes
\langle \phi |,  \label{F11}
\end{eqnarray}
and 
\begin{eqnarray}
{\widetilde O}|\Psi \rangle &{\rightarrow}&[1\otimes O]|\Psi \rangle 
\nonumber \\
&=&[1\otimes O]|\phi \rangle \otimes \langle \phi |=|\phi \rangle \otimes
\langle \phi |O^{\dagger }.  \label{F12}
\end{eqnarray}

As a consequence, we show that the invariant $(P^2)^{\widehat{}}$ has a null
constant value in this representation, that is 
\begin{equation}
(P^2)^{\widehat{}}\ \ |\Psi \rangle =(2P^\mu \widehat{P}_\mu -\widehat{P}%
^\mu \widehat{P}_\mu )|\Psi \rangle =0.  \label{p46}
\end{equation}
This is but the density matrix equation for the Klein-Gordan field derived
in ${\it I}$.

In order to construct a spinor density matrix equation, we introduce an
equation as 
\begin{equation}
(\alpha ^\mu {P}_\mu )^{\widehat{}}\ \ |\Psi \rangle =0,  \label{d47}
\end{equation}
such that, differently from Eqs.(\ref{p46}), we write

\begin{equation}  \label{d48}
(\alpha ^{\mu}{P}_{\mu})^{\widehat{}}\ \ (\alpha ^{\mu}{P}_{\mu})^{\widehat{}%
}=(P^2)^{\widehat{}}.
\end{equation}
Using Eq. (\ref{dois}), a generic solution is immediately found to be 
\begin{equation}  \label{d49}
\alpha ^{\mu}=\sigma \gamma ^{\mu},
\end{equation}
where $\gamma ^{\mu}$ are the Dirac matrix, and $\sigma $ is some non null
Lorentz invariant. A trivial choice is $\sigma =1$ (but another possibility
for $\sigma $ is explored in the next section). Therefore we find 
\begin{eqnarray}
\left( \gamma ^{\mu}{P}_{\mu} \right)^{\widehat{}}\ \ |\Psi \rangle &=&
[(\gamma ^{\mu}{P}_{\mu})\otimes 1 - 1\otimes (\gamma ^{\mu}{P}_{\mu})]|\phi
\rangle \otimes \langle \bar{\phi} |  \nonumber \\
&=&(\gamma ^{\mu}{P}_{\mu})|\phi \rangle \otimes \langle \bar{\phi} |- |\phi
\rangle \otimes \langle \bar{\phi} |(\gamma ^{\mu}{P}_{\mu})^{\dagger}=0.
\label{d51}
\end{eqnarray}
where now $|\Psi>$ is a $16$-component spinor, and $|\phi \rangle (\langle 
\bar{\phi}|)$ is the $4$- (dual) Dirac spinor.

Multiplying the rhs of Eq.(\ref{d51}) by $|\phi \rangle$, it results in $%
(\gamma ^{\mu }{p}_{\mu }-m)\ |\phi \rangle =0 $, the Dirac equation. Now,
multiplying the lhs of Eq.(\ref{d51}) by $\langle \bar{\phi} |$ it results
in $\langle \bar{\phi} |({p}_{\mu }\gamma ^{\mu }-m) =0 $, the conjugated
Dirac equation. In this sense, in fact, Eq.(\ref{d51}) is a Liouville-von
Neumann-like equation for the Dirac Field.

At this point, we are able to explore some symmetry properties of Eq. (\ref
{d51}), as the invariance under similarity transformations. That is,
consider 
\begin{equation}
\widehat{\left( \gamma ^{\mu }P_{\mu } \right)^{\prime }}\ =U\ \widehat{%
\left( \gamma ^{\mu}P_{\mu } \right)}\ \ U^{-1},  \label{d71}
\end{equation}
and 
\begin{equation}
|\Psi \rangle ^{\prime }=U|\Psi \rangle .  \label{d72}
\end{equation}
Then Eq.(\ref{d51}) reads 
\begin{equation}  \label{61}
\widehat{\left( \gamma ^{\mu }P_{\mu } \right)^{\prime }}\ |\Psi \rangle
^{\prime}=0.
\end{equation}
On the other hand, if $[U, ( \gamma ^{\mu }{P}_{\mu } )^{\widehat{}}\ \ ]=0$
then $|\Psi \rangle^{^{\prime}}$ given in Eq.(\ref{d72}) is a solution of
Eq.(\ref{d51}). In this case, one suggestive example is provided by $U=U(
y^{\mu }{\widehat{P}}_{\mu } )$ written in the form 
\begin{equation}  \label{u33}
U=U( y{\widehat{P}})=\exp [-i y{\widehat{P}}],
\end{equation}
where $y$ is the transformation parameter. Accordingly, Eq.(\ref{d72}) reads

\[
|\Psi (y)\rangle ^{^{\prime }}=\exp (-iyP)|\phi \rangle \otimes \langle \bar 
\phi |\exp (iyP^{\dagger }). 
\]

The gauge covariance can be considered if we write 
\begin{equation}
P\ \ \rightarrow \ \ -id_\mu =P_\mu +gA_\mu ,  \nonumber
\end{equation}
where $d_\mu $ is the usual covariant derivative. Then 
\begin{equation}
|\Psi (y)\rangle ^{^{\prime }}=\exp (-yd)|\phi \rangle \otimes \langle \bar 
\phi |\exp (yd^{\dagger }).  \label{gaugeh1}
\end{equation}
where the operation $\otimes $ is defined by Eq.(\ref{est11}). This function 
$|\Psi (y)\rangle $ is a solution of the density matrix equation, which is
derived from Eq.(\ref{d51}), but rather considering the gauge field, that is

\begin{equation}
\lbrack (\gamma ^\mu {d}_\mu )\otimes 1-1\otimes (\gamma ^\mu {d}_\mu
)]|\phi \rangle \otimes \langle \bar \phi |=0.  \label{d52}
\end{equation}
If we use the definition given by Eq.(\ref{est11}), then we can write $%
\langle x,x^{\prime }|\Psi \rangle =\Psi (x,x^{\prime })$, such that Eq.(\ref
{gaugeh1}) can be written as

\begin{equation}
\Psi (x,y)=\exp {[-yd(x)]}\ \phi (x)\otimes \bar \phi (x)\ \exp {%
[yd(x)^{\dagger }]}.  \label{gaugeh33}
\end{equation}
which is the generalized Heinz density operator. It is to be emphasized that 
$|\phi \rangle $ (or $\phi (x)$) are elements of the Hilbert space. So the
various quantum numbers either from the space-time group or from the gauge
group are assumed to be attached to each element of the representation space.

As another example of the use of such symmetry approach, we derive the TFD
spinor-electrodynamics equations. The fields are assumed to be the basic
dynamical variables, so that the Lagrangian may be written as

\begin{eqnarray*}
L &=&\bar \phi \widehat{D}\phi +\bar \phi (D-\widehat{D}-m)\widehat{\phi } \\
&&+\widehat{\bar \phi }(D-\widehat{D}-m)\phi -\widehat{\bar \phi }(D-%
\widehat{D}-m)\widehat{\phi } \\
&&+{\frac 12}F^{\mu \nu }{\widehat{F}}_{\mu \nu }-{\frac 14}{\widehat{F}}%
^{\mu \nu }{\widehat{F}}_{\mu \nu },
\end{eqnarray*}
which is derived from the usual spinor-electrodynamics Lagrangian and the
hat mapping, and where

\[
D=i\gamma ^\mu ({\partial }_\mu +ieA_\mu ). 
\]
Therefore, it follows that

\begin{eqnarray*}
{\widehat{D}}\phi +(D+m)\widehat{\phi }-{\widehat{D}}\widehat{\phi } &=&0, \\
(D+m)\phi &=&0,
\end{eqnarray*}
which are the quasi-particle TFD equations for the spinor field. The other
equations can be derived accordingly.

In the next section we are going to present another application of the
method developed here by deriving a phase space Wigner structure attached to
the Dirac field. We use the more general form of Eq.(\ref{d51}) including $%
\sigma \neq 1$, in such a way that we can better compare our formalism with
the one proposed by Bohm and Hiley\cite{bo1} and Holland\cite{bo2} to derive
phase-space spinor fields under a geometric perspective.

\section{The Dirac field and a collisionless Boltzmann equation}

Notice that, using Eq.(\ref{d49}), Eq.(\ref{d47}) can be written as 
\begin{equation}
(\sigma ^l\gamma ^{l\mu }\frac \partial {\partial x^\mu }-\sigma ^r\gamma
^{r\mu }\frac \partial {\partial x^{\prime \mu }})\Psi _{ts}(x,x^{\prime
})=0,  \label{dir1}
\end{equation}
such the $\Psi _{ts}(x,x^{\prime })=\phi (x)_t{\bar \phi }_s(x^{\prime })$,
where the subscripts $t,s=1,2,..,4$ refer to the two spinor indices.

Here $\gamma ^{l\mu }$ $=\gamma ^\mu \otimes 1$, $\gamma ^{r\mu }$ $%
=1\otimes \gamma ^\mu $ are $16\times 16$ matrices and $\gamma ^\mu $ are
the Dirac matrices; $\sigma ^l=\sigma \otimes 1$ and $\sigma ^r$ $=1\otimes
\sigma ;$ with $\sigma $, the arbitrary Lorentz invariant to be specified;
besides the $\gamma ^{r,l}$-matrices fulfil two Clifford algebras: 
\begin{equation}
\{\gamma ^{l\mu },\gamma ^{l\nu }\}=\{\gamma ^{r\mu },\gamma ^{r\nu
}\}=2g^{\mu \nu };\,\{\gamma ^{l\mu },\gamma ^{r\nu }\}=0  \label{cliff1}
\end{equation}

Multiplying Eq.(\ref{dir1}) by $\left( \sigma ^l\right) ^{-1}$, it follows
that 
\begin{equation}
\Lambda (\kappa )\Psi _{ts}(x,x^{\prime })=(\gamma ^{l\mu }\frac \partial {%
\partial x^\mu }-\kappa \gamma ^{r\mu }\frac \partial {\partial x^{\prime
\mu }})\Psi _{ts}(x,x^{\prime })=0,  \label{dirac3}
\end{equation}
where $\kappa =(\sigma ^l)^{-1}\sigma ^r$. Using the following
transformation 
\begin{equation}
\frac \partial {\partial x^\mu }=\frac 1{\sqrt{2}}(\frac \partial {\partial
q^\mu }-p_\mu ),\ \ \ \frac \partial {\partial \ x^{\prime \mu }}=\frac 1{%
\sqrt{2}}(\frac \partial {\partial q^\mu }+p_\mu ),  \label{oper21}
\end{equation}
the square of $\Lambda (\kappa )$ can be written as 
\begin{equation}
\Lambda ^2(\kappa )=\Lambda _1^{\mu \nu }(\kappa )\frac{\partial ^2}{%
\partial q^\mu \partial q^\nu }+\Lambda _2^{\mu \nu }(\kappa )\frac \partial
{\partial q^\mu }\,p_\nu +\Lambda _3^{\mu \nu }(\kappa )\,p_\mu \frac 
\partial {\partial q^\mu }+\Lambda _2^{\mu \nu }(\kappa )p_\mu p_\nu ,
\label{lamb1}
\end{equation}
where 
\begin{eqnarray*}
\Lambda _1^{\mu \nu }(\kappa ) &=&\frac 12(\gamma ^{l\mu }-\kappa \gamma
^{r\mu })(\gamma ^{l\nu }-\kappa \gamma ^{r\nu }), \\
\Lambda _2^{\mu \nu }(\kappa ) &=&\frac 12(\gamma ^{l\mu }+\kappa \gamma
^{r\mu })(\gamma ^{l\nu }-\kappa \gamma ^{r\nu }), \\
\Lambda _3^{\mu \nu }(\kappa ) &=&\frac 12(\gamma ^{l\mu }-\kappa \gamma
^{r\mu })(\gamma ^{l\nu }+\kappa \gamma ^{r\nu }), \\
\Lambda _4^{\mu \nu }(\kappa ) &=&\frac 12(\gamma ^{l\mu }+\kappa \gamma
^{r\mu })(\gamma ^{l\nu }+\kappa \gamma ^{r\nu }).
\end{eqnarray*}

{}From these expressions for $\Lambda _i^{\mu \nu }(\kappa ),\,(i=1,..,4)$,
the following operators can be defined 
\begin{equation}
a^{\mu +}(\kappa )=\frac 1{\sqrt{2}}(\gamma ^{l\mu }+\kappa \gamma ^{r\mu
})\,\,\,\, a^\mu (\kappa )=\frac 1{\sqrt{2}}(\gamma ^{l\mu }-\kappa \gamma
^{r\mu });  \label{aoper1}
\end{equation}
then Eq.(\ref{lamb1}) is rewritten as 
\begin{eqnarray*}
\Lambda ^2(\kappa ) &=&a^\mu (\kappa )a^\nu (\kappa )\frac{\partial ^2}{%
\partial q^\mu \partial q^\nu }+a^{\mu +}(\kappa )a^\nu (\kappa )\frac 
\partial {\partial q^\mu }\,\,p_\nu \\
&&+a^\mu (\kappa )a^{\nu +}(\kappa )\,\,p_\mu \,\frac \partial {\partial
q^\mu }+a^{\mu +}(\kappa )a^{\nu +}(\kappa )p_\mu p_\nu .
\end{eqnarray*}

Notice that if $\kappa $ anticommutes with the matrices $\gamma ^{l\mu }$
and $\gamma ^{r\mu }$, and satisfies $\kappa ^2=1$ (which is compatible with
Eq.(\ref{d51})), it leads to 
\begin{eqnarray*}
\{a^\mu ,a^\nu \} &=&\{a^{\mu +},a^{\nu +}\}=0 \\
\{a^{\mu +},a^\nu \} &=&g^{\mu \nu },
\end{eqnarray*}

Observe that the requirement of Lorentz invariance for $\kappa $ can be
achieved if we define $\sigma $ in Eq.(\ref{d49}) as $\sigma =\gamma ^5$
such that $\kappa =\gamma ^{l5}\gamma ^{r5}$, where $\gamma ^5=i\gamma
^1\gamma ^2\gamma ^3\gamma ^4$. Using these results, we follow Bohm and Hiley%
\cite{bo1} to write 
\begin{eqnarray*}
\Lambda ^2 &=&a^{\mu +}a^\nu \frac \partial {\partial q^\mu }\,\,p_\nu
+a^\mu a^{\nu +}\,\,p_\mu \,\frac \partial {\partial q^\mu } \\
\ &=&p^\mu \frac \partial {\partial q^\mu }.
\end{eqnarray*}
Therefore, from Eq.(\ref{dirac3}), we obtain 
\begin{equation}
p^\mu \frac \partial {\partial q^\mu }\,\Psi (q,p)=0;  \label{wigr1}
\end{equation}
which is the Bohm and Hiley results\cite{bo1}, as developed by Holland\cite
{bo2}, about the relativistic phase space. But here, we have derived Eq.(\ref
{wigr1}) from first principles, so that we can go somewhat further with it.
Observe, in particular, that Eq.(\ref{wigr1}) can be derived also for the
Klein-Gordon field if we use the transformation given by Eq.(\ref{oper21})
in Eq.(\ref{p46}). It should be stressed here that within an approach based
entirely on representations of symmetry, only a collisionless Boltzmann
equation can be obtained. The role of dynamics can be included by
postulating the interaction Lagrangian, that will lead to collision terms.
The result, then, will be model-dependant. Consistently, Eq.(\ref{wigr1})
describes only a drift term, that should be the same for any kind of
systems, including the classical ones.

In particular, in this formalism we take advantage of the fact that the
state is treated as an amplitude, to implement higher order terms in Eq.(\ref
{wigr1}) looking for an invariant Lagrangian procedure, avoiding then the
(usual) intricate way to do that. Indeed, observe that one simple way to
derive a Vlasov term (the influence of a external field via $F^\mu )$ in Eq.(%
\ref{wigr1}), considering the first order (classical) limit in the Heinz
approach, is to write the Lorentz invariant Lagrangian as

\begin{eqnarray}
{\sc L}(q,p) &=&{\frac 12}\Psi ^{*}(q,p)p_\mu {\partial }^\mu \Psi (q,p)-{%
\frac 12}\Psi (q,p)p_\mu {\partial }^\mu \Psi ^{*}(q,p)  \nonumber \\
&&-{\frac 12}\Psi ^{*}(q,p)mF_\mu {\partial }_{(p)}^\mu \Psi (q,p)-{\frac 12}%
\Psi (q,p)mF^\mu {\partial }_{(p)}^\mu \Psi ^{*}(q,p).  \label{vlas11}
\end{eqnarray}
Then the energy-momentum tensor is defined as

\begin{equation}
T^{\mu \nu }=\int d^3p\frac 1{p^o}\Psi ^{*}(q,p)p^\mu p^\nu \Psi (q,p),
\label{tens12}
\end{equation}
which is still compatible with the usual definition of $T^{\mu \nu }$ since $%
\Psi ^{*}(q,p)\Psi (q,p)$ can be interpreted as the classical distribution
function in the relativistic phase space.

Another aspect to be explored in the present formalism is the inclusion of
gravitational effects in the kinetic theory. The treatment of transport
theory in a curved space-time background is somewhat hindered by the fact
that a definition of the Wigner function {\em $\Psi (q,p){\ }$}depends on
the use of the Fourier transform of a spacetime correlation function with a
translated argument. Fourier transformation (nor translations) are globally
available in general curved spacetime. However, as shown by Calzetta, Habib
and Hu\cite{ca1} , the construction is still meaningful, if carried out in
the tangent space of an event in spacetime and related to the neighbourhood
of that event by an exponential map. Therefore, starting from Eq.(\ref{d52}%
), the Liouville-Vlasov equation for the Wigner function of a spinor field
coupled to a gauge field with field strength tensor ${F}^{\mu \nu }$ is then%
{\em \ 
\[
(-{p}_\mu {F}^{\mu \nu }{D}_\nu +{p}^\mu {D}_\mu )\Psi (q,p)=0, 
\]
}where ${D}_\mu $ is the covariant derivative including the curved
space-time spin connection.

\section{Final conclusions and remarks}

Summarizing, we have used here a representation formalism for Lie algebras,
in which the elements of the representation space, say ${\cal H}_w$, are
closely connected with the notion of density matrix through the thermofield
dynamics approach. We have started with the Poicar\'e Lie algebra in the $%
{\cal H}_w$-space. Then we have derived a gauge invariant density matrix
equation for the spinorial field, and shown that one particular solution for
such an equation is the gauge invariant density operator as proposed by
Heinz. This fact has set forth a solid structure to support the Heinz
approach, since it has been derived here by an analysis of group theory,
i.e. the fundamental symmetry properties of the system. In the same context,
analysing a Lagrangian formalism, we have derived the TFD equations; and so,
we have presented an association of the Heinz density method and TFD. The
phase space structure of the spinorial field has also been studied as
another application of our method. In this case, once again, we have been
able to advocate the group theoretical bases to derive the early results of
Bohm and Hiley\cite{bo1} and Holland\cite{bo2} about the connection of the
Clifford algebra and phase-space. But here we have proceeded somewhat
further, pointing out a possible Lagrangian formalism to take into account
higher order terms in a kinetic equation. This last fact opens the
possibility to deal with the kinetic theory as a genuine quantum or
classical field formalism (observe that, as we are analysing representations
of a Lie algebra, the function $\Psi (p,q)$ in Eq.(\ref{wigr1}) can be
considered as a quantum-like field defined in a Fock space, similar to what
has been developed for Galilean classical systems\cite{ak3,ak4}); and so, a
pertubative scheme can be introduced by symmetry properties of the
Lagrangian, as has been argued in writing Eq.(\ref{vlas11}). Besides, we
have discussed the effect of the curved space-time background for the
spinorial density matrix equation. Some of these aspects are treated in a
more detailed way in another paper\cite{ak8}).

%
%

{\bf Acknowledgements:}

This work is supported in part by the Natural Sciences and Engineering
Research Council of Canada and by CNPq and CAPES (two Brazilian Government
Agencies for Research). AES thanks the Theoretical Physics Institute for the
hospitality during his visit to Edmonton.

\end{document}